\documentclass[12pt,notitlepage]{article}
\usepackage{graphics}
\usepackage{amsmath,amssymb}
\usepackage{graphicx}
\usepackage{cite}

\usepackage{color}

\def\ignore#1{{}}

\newcounter{sxn}

\newcounter{axn}

\newdimen\mybaselineskip
\newcommand{\beeq}{\begin{equation}}
\newcommand{\eneq}{\end{equation}}
\newcommand{\beqn}{\begin{eqnarray}}
\newcommand{\eeqn}{\end{eqnarray}}

\newcommand{\ba}{\begin{array}}
\newcommand{\ea}{\end{array}}

\newcommand{\be}{\begin{equation}}
\newcommand{\ee}{\end{equation}}
\newcommand{\bea}{\begin{eqnarray}}
\newcommand{\eea}{\end{eqnarray}}
\newcommand{\beal}{\setcounter{letter}{1} \begin{eqnarray}}
\newcommand{\eeal}{\addtocounter{equation}{1} \end{eqnarray}}

\newcommand{\larrow}{\,\,\,\,\hbox to 30pt{\rightarrowfill}
\,\,\,\,}
\newcommand{\slarrow}{\,\,\,\hbox to 20pt{\rightarrowfill}
\,\,\,}

\def\la{\raise.16ex\hbox{$\langle$}\lower.16ex\hbox{}  }
\def\ra{\, \raise.16ex\hbox{$\rangle$}\lower.16ex\hbox{} }

\def\psibar{ \psi \kern-.65em\raise.6em\hbox{$-$} \lower.6em\hbox{} }
\def\psibarb{ \psi \kern-.65em\raise.6em\hbox{$-$}  }

\begin{document}

\thispagestyle{empty}





\begin{center}  
{\LARGE \bf  Gravitational and electromagnetic radiation from an electrically charged black hole in general nonlinear electrodynamics}

\vspace{1cm}

{\bf  Ramin G.~Daghigh$^{1}$ and Michael D.\ Green$^2$}
\end{center}

\centerline{\small \it $^1$ Natural Sciences Department, Metropolitan State University, Saint Paul, Minnesota, USA 55106}
\vskip 0 cm
\centerline{} 

\centerline{\small \it $^2$ Mathematics and Statistics Department, Metropolitan State University, Saint Paul, Minnesota, USA 55106}
\vskip 0 cm
\centerline{} 

\vspace{1cm}

\begin{abstract}
We derive the equations for the odd and even parity perturbations of coupled electromagnetic and gravitational fields of a black hole with an electric charge within the context of general nonlinear electrodynamics.  The Lagrangian density is a generic function of the Lorentz invariant scalar quantities of the electromagnetic fields.   We include the Hodge dual of the electromagnetic field tensor and the cosmological constant in our calculations.  For each type of parity, we reduce the system of Einstein field equations coupled to nonlinear electrodynamics to two coupled Schr\"odinger-type wave equations, one for the gravitational field and one for the electromagnetic field.   The stability conditions in the presence of the Hodge dual of the electromagnetic field are derived.
\end{abstract}

\newpage

\section{Introduction}

Penrose, in his Noble prize winning work\cite{Penrose1965}, shows that when a massive star collapses to form a black hole, the singularity formation   in general relativity (GR) is inevitable. This issue signals the demise of GR in its classical form.   
The singularity may be resolved by an ultimate quantum theory of gravity that can describe the final stage of gravitational collapse.  In the absence of a microscopic theory, toy models of  regular (singularity-free) black holes have been proposed to 
study the formation and evaporation of such black holes.  After the first specific proposal
for a regular black hole (RBH), which was presented by Bardeen in \cite{reg1}, many RBH models have been proposed by various authors over the years.  See, for example, \cite{PoissonIsrael, reg2,reg3,reg4, reg5, reg6, reg7, reg9, reg10,reg11, reg12, reg13, reg14, reg15} for some of the RBHs that are asymptotically Schwarzschild at large radii.  The majority of these black holes, including Bardeen's model, are constructed in an ad-hoc manner without an underlying theory behind them.  However, in \cite{Ayon1}, Ay\'on-Beato and Garc\'ia  found the first RBH solution in GR that is coupled to nonlinear electrodynamics (NLED). NLED was originally proposed in \cite{Born}, by Born and Infeld, in an attempt to generalize Maxwell's theory to strong field regimes.  As a result, this theory provides a natural choice for studying charged black holes where we deal with strong electromagnetic and gravitational fields.

Ay\'on-Beato and Garc\'ia were also able to reinterpret Bardeen's model as a  black hole with a nonlinear magnetic monopole charge in \cite{Ayon2}.  It was also shown by Rodrigues and Silva in \cite{Silva} that the Bardeen solution can be obtained in NLED with an electric charge.  In addition to the electrically charged black hole in \cite{Ayon1},   Ay\'on-Beato and Garc\'ia proposed two more black hole models with electric charge in  \cite{Ayon1a} and \cite{Ayon1b}.

For these RBH models to be viable, they need to be stable when they are perturbed.  In addition to its relevance to gravitational wave observations, the study of black hole perturbations is crucial in determining the stability of a black hole model \cite{RW}. There are two approaches to study black hole perturbations.  In one approach, the perturbation of a field (e.g.\ a scalar field), which is weakly coupled to the background of a black hole spacetime, is analyzed.  In this case, the geometric perturbations are usually neglected.  Since the equations governing the perturbations of spherically symmetric black holes are similar to the Klein-Gordon equation for a scalar field, one can achieve a qualitative understanding of how the RBH and its perturbations differ from its Schwarzschild counterpart.  However, to achieve a quantitative understanding of the stability of a black hole, one needs to look at the perturbations of the spacetime and any strongly coupled fields to the background geometry.  

The wave equations of coupled electromagnetic and gravitational fields of a black hole with an electric charge in general NLED are derived for the first time by Moreno and Sarbach in \cite{Moreno}.     The Lagrangian considered in \cite{Moreno} is a general function of the Lorentz invariant scalar quantity $F$ of the electromagnetic field, where $F=\frac{1}{4}F_{\mu \nu} F^{\mu \nu}$ and $F_{\mu \nu}$ is the electromagnetic field tensor. The stability conditions for these black holes are also derived in \cite{Moreno}.

The wave equations of coupled electromagnetic and gravitational fields of a black hole with a magnetic monopole charge in general NLED are derived for the first time by Nomura {\it et al.} in \cite{Nomura}.   In addition to the electromagnetic field tensor $F_{\mu \nu}$,  the authors of \cite{Nomura} include the Hodge dual of the field tensor, $F_{*\mu \nu}$.  

Since magnetic monopoles have never been observed in nature, in this paper we focus on electrically charged black holes within the context of  NLED.  Similar to \cite{Moreno}, we introduce perturbations on the background geometry of a charged black hole and its nonlinear electric field.  We derive the odd parity (magnetic or axial) and even parity (electric or polar) wave equations for the coupled electromagnetic and gravitational fields. In our calculations, we include the Hodge dual field tensor, which was ignored in \cite{Moreno}.  The method we use in this paper is different than the gauge-invariant approach used in \cite{Moreno}.  Our method, where we fix the gauge (i.e.\ Regge-Wheeler gauge) early on, is more in line with the work done by Zerilli in \cite{Zerilli} for the Reissner-Nordstr\"om black hole and by Nomura {\it et al.} in \cite{Nomura} for black holes with a magnetic monopole.
For simplicity, we do not consider any test particle outside the black hole horizon.  However, it should be easy to incorporate that using Zerilli's results in \cite{Zerilli}.

We structure the paper as follows. In Section \ref{Sec:WE}, we set up the problem by deriving the perturbed Einstein-NLED equations.  In Section \ref{Sec:opp}, we expand the geometric and NLED perturbations in tensor harmonics and derive the wave equations for odd parity perturbations, which are reduced to two coupled Schr\"odinger-type wave equations.  
We then derive stability conditions in Section \ref{Sec:SOdd}.
In Section \ref{Sec:epp}, we examine the even parity perturbations.   In Section \ref{Sec:application}, to provide an example of a theory with a Hodge dual field, we apply our stability conditions to RBHs in Einstein-Born-Infeld gravity.   We provide the summary and conclusion in Section \ref{Sec:conclusions}.  In Appendices \ref{Sec:WEEven} and \ref{Sec:SEven}, we provide more details and calculations involving even parity perturbations and their stability. In Appendix C, we explain some of the differences between our results, when reduced to the Reissner-Nordstr\"om case, and Zerilli's results in \cite{Zerilli}.

\section{Perturbed Field Equations}
\label{Sec:WE}

In order to make the comparison with the Reissner-Nordstr\"om black hole perturbations easier, we closely follow the notation in \cite{Zerilli}.  The action of NLED in a curved spacetime is
\begin{equation}
	\mathcal{S}=  \int d^4x \sqrt{-g}\left( \frac{1}{16\pi} (R-2\Lambda)-\frac{1}{4\pi}\mathcal{L}(F,F_*)\right),
	\label{}
\end{equation}
where $R$ is the Ricci scalar, $\Lambda$ is the cosmological constant, $g$ is the determinant of the spacetime metric tensor $g_{\mu \nu}$, and the Lagrangian density $\mathcal{L}$ is an arbitrary function of the Lorentz invariant scalar quantities\footnote{Note that $\Lambda$ can be absorbed by a redefinition of the Lagrangian density.  However, since it does not add much complexity, we will keep it throughout our calculations.}
\begin{equation}
	F=  \frac{1}{4} F_{\mu \nu}  F^{\mu \nu}
	\label{}
\end{equation}
\begin{equation}
	F_*=  \frac{1}{4} F_{\mu \nu}  F_*^{\mu \nu}.
	\label{}
\end{equation}
Here, $F_*^{\mu \nu}=\frac{1}{2} \epsilon^{\mu \nu \alpha \beta} F_{\alpha \beta}$ is the Hodge dual of the the electromagnetic field tensor $F^{\mu \nu}$.  The Levi-Civita tensor is normalized as $\epsilon_{0123}=\sqrt{-g}$.  In this paper, we adopt Planck units where $c=G=\hbar=1$. The Einstein-NLED equations that describe the gravitational and NLED fields are
\begin{equation}
	\tilde{G}_{\mu \nu}=8\pi \tilde{E}_{\mu \nu}
	\label{EinsteinEq}
\end{equation}
\begin{equation}
	\left(\sqrt{-\tilde{g}} \tilde{\mathcal{L}}_{\tilde{F}}  \tilde{F}^{\mu \nu}  + \sqrt{-\tilde{g}} \tilde{\mathcal{L}}_{\tilde{F}_*}  \tilde{F}_*^{\mu \nu}\right)_{, ~\nu}=0,
	\label{EM-Eq}
\end{equation}
where  $\tilde{\mathcal{L}} \equiv \mathcal{L}(\tilde{F}, \tilde{F_*})$, $\tilde{\mathcal{L}}_{\tilde{F}} \equiv \partial  \tilde{\mathcal{L}} / \partial \tilde{F}$, and $\tilde{\mathcal{L}}_{\tilde{F_*}} \equiv \partial  \tilde{\mathcal{L}} / \partial \tilde{F_*}$.  We use tilde for quantities associated with the {\it total} NLED and gravitational fields. Quantities with no tilde refer to the background geometry represented by the static spherically symmetric line element\footnote{Note that we assume $g_{rr}=-1/g_{tt}$.  It turns out this is forced to be true.  Had we not made this assumption, once we get to Eq.\ (\ref{EinsteinEqRaised}), using $G_r{}^r-G_t{}^t=0$ we find $-g_{tt} g_{rr}$ is a constant.  The constant can be set to one by rescaling the time coordinate.  For more details, see \cite{Nomura}.}
\begin{equation}
	ds^2=-e^\nu dt^2 +e^{-\nu} dr^2 + r^2(d\theta^2+\sin^2 \theta d\phi^2).
	\label{LineElement}
\end{equation}

We assume the following general ansatz for the Maxwell field for an electric charge
\begin{equation}
	F_{\mu \nu }=2\delta^t_{[\mu}\delta^r_{\nu]} B(r,\theta, \phi)
	\label{Fmn}
\end{equation}
from which we get $F_*=0$.  One then can integrate Eq.\ (\ref{EM-Eq}) to obtain
\begin{equation}
	F_{\mu \nu }=2\delta^t_{[\mu}\delta^r_{\nu]} \frac{f(\theta, \phi)}{r^2 \mathcal{L}_{F}}.
	\label{}
\end{equation}
In the case of spherical symmetry, the invariant quantities $F$ and $F_*$ only depend on the radial coordinate.  Consequently, both $\mathcal{L}$ and $\mathcal{L}_F$ are functions of the radial coordinate only.  Therefore, one can use the Bianchi identity, $d\textbf{F}=0$, to show  $f(\theta, \phi)$ is a constant that we will call $-q$.  Therefore, the background field strength can be written as
\begin{equation}
	{\bf{F}}= -\frac{q}{r^2 \mathcal{L}_{F}} dt \wedge dr.
	\label{Bianchi}
\end{equation}
In the rest of this paper, we choose
\begin{equation}
	Q(r)= \frac{q}{ \mathcal{L}_{F}}.
	\label{}
\end{equation}
In the case of the Reissner-Nordstr\"{o}m black hole, where $\mathcal{L}_{F}=1$, $q$ is simply the electric charge.  In the RBH models presented in \cite{Ayon1, Ayon1a, Ayon1b}, $q$ is also interpreted as the electric charge.  
Equation (\ref{Bianchi}) gives us the background values of the invariant scalar quantities as 
\begin{equation}
	F=-\frac{Q^2}{2r^4},~~~~F_*=0.
	\label{ScalarF}
\end{equation} 
In the right-hand-side of Eq.\ (\ref{EinsteinEq}), we have\cite{Nomura}
\begin{equation}
	\tilde{E}_{\mu \nu}=\frac{1}{4\pi} \left[ \tilde{\mathcal{L}}_{\tilde{F}} \tilde{g}^{\rho \sigma} \tilde{F}_{\rho \mu }\tilde{F}_{\sigma \nu} + \tilde{g}_{\mu \nu}\left(\tilde{\mathcal{L}}_{\tilde{F}_*}\tilde{F}_*-\tilde{\mathcal{L}} -\frac{\Lambda}{2}\right) \right].
	\label{Emn}
\end{equation}
Note that in the case of the Reissner-Nordstr\"om black hole, $\Lambda=0$ and $\tilde{\mathcal{L}}=\tilde{F}$.

We introduce first-order perturbations $\tilde{g}_{\mu \nu} = g_{\mu \nu}+ h_{\mu \nu}$ and $\tilde{F}_{\mu \nu}= F_{\mu \nu} + f_{\mu \nu}$ assuming $|h_{\mu \nu}| \ll 1$ and $|f_{\mu \nu}| \ll 1$. Also, to first-order, $\tilde{g}^{\mu \nu} = g^{\mu \nu}- h^{\mu \nu}$, where $h^{\mu \nu}=g^{\mu \alpha}g^{\nu \beta} h_{\alpha \beta}$, and $\sqrt{-\tilde{g}}=\sqrt{-g}(1+\frac{1}{2}g^{\mu \nu}h_{\mu \nu})$. Substituting these into Eqs.\ (\ref{EinsteinEq}) and (\ref{EM-Eq}) and keeping terms to first order, we arrive at the perturbed Einstein-NLED equations
\begin{equation}
	\delta G_{\mu \nu}=8\pi \delta E_{\mu \nu}
	\label{}
\end{equation}
\begin{equation}
	\delta \left(\sqrt{-\tilde{g}} \tilde{\mathcal{L}}_{\tilde{F}}  \tilde{F}^{\mu \nu}  + \sqrt{-\tilde{g}} \tilde{\mathcal{L}}_{\tilde{F}_*}  \tilde{F}_*^{\mu \nu}\right)_{, ~\nu}=0,
	\label{}
\end{equation}
or more specifically\footnote{We correct a typo in \cite{Zerilli}, which appears in the fourth term of the left-hand-side of Eq.\ (\ref{Einstein-h}).  The correct expression can also be found in \cite{Beetle}.}
\begin{eqnarray}
	&&h_{\mu \nu ;\alpha}{}^{;\alpha}- (h_{\mu \alpha}{}^{;\alpha}{}_{;\nu}+h_{\nu \alpha}{}^{;\alpha}{}_{;\mu})+2 R_\mu{}^{\alpha}{}_{\nu}{}^{\beta} h_{\alpha \beta} + h^{\alpha}{}_{\alpha} {}_{;\mu}{}_{;\nu} \nonumber \\
	&& -(R^\alpha{}_\nu h_{\mu \alpha}+ R^\alpha{}_\mu h_{\nu \alpha}) + R h_{\mu \nu} \nonumber \\
	&& + g_{\mu \nu}(h_{\alpha \beta}{}^{;\beta;\alpha} - h^\alpha {}_{\alpha; \beta}{}^{;\beta}- R^{\alpha \beta} h_{\alpha \beta}) = -16 \pi \delta E_{\mu \nu}
	\label{Einstein-h}
\end{eqnarray}
and
\begin{eqnarray}
	&&\left\{ \sqrt{-g} \left[ \mathcal{L}_{F}  f^{\mu \nu}  + \mathcal{L}_{F_*}  f_*^{\mu \nu}   +  \left( \mathcal{L}_{FF} \delta F^{(f)} + \mathcal{L}_{FF_*} \delta F_*^{(f)} \right) F^{\mu \nu}+ 
	 \left( \mathcal{L}_{F_*F_*} \delta F_*^{(f)} + \mathcal{L}_{F_*F} \delta F^{(f)} \right)F_*^{\mu \nu}  \right] \right\}_{, ~\nu}  \nonumber \\
	 &&=\left[ \sqrt{-g}\mathcal{L}_{F} \left( h^{\mu \alpha} g^{\nu \beta}+g^{\mu \alpha} h^{\nu \beta} \right) F_{\alpha \beta} - \frac{1}{2} \sqrt{-g}  \mathcal{L}_{F} F^{\mu \nu}g^{\alpha \beta} h_{\alpha \beta} \right. \nonumber \\
	&&  - \sqrt{-g} \left( \mathcal{L}_{FF} \delta F^{(h)} + \mathcal{L}_{FF_*} \delta F_*^{(h)} \right)F^{\mu \nu} - 
	\sqrt{-g}  \left( \mathcal{L}_{F_*F_*} \delta F_*^{(h)} + \mathcal{L}_{F_*F} \delta F^{(h)}\right) F_*^{\mu \nu} \bigg]_{, ~\nu}
	\label{EM-f}
\end{eqnarray}
where
\begin{equation}
	\delta E_{\mu \nu} = \delta E^{(h)}_{\mu \nu} + \delta E^{(f)}_{\mu \nu}
	\label{}
\end{equation}
in which
\begin{eqnarray}
	\delta E^{(h)}_{\mu \nu} &=& -\frac{1}{4\pi}  \left\{  \mathcal{L}_F F_{\alpha \mu}F_{\beta \nu} h^{\alpha \beta}  - \left(\mathcal{L}_{F_*} F_* -\mathcal{L} - \frac{\Lambda}{2}\right) h_{\mu \nu}   \right. \nonumber \\
	&& \left. +\left( \mathcal{L}_F g_{\mu \nu}  - \mathcal{L}_{FF} g^{\alpha \beta} F_{\alpha \mu} F_{\beta \nu}   -  \mathcal{L}_{F_*F} F_* g_{\mu \nu} \right) \delta F^{(h)}  \right. \nonumber \\
	&&  -\left(  \mathcal{L}_{FF_*} g^{\alpha \beta} F_{\alpha \mu} F_{\beta \nu}   + \mathcal{L}_{F_*F_*} F_* g_{\mu \nu} \right) \delta F_*^{(h)}  \bigg\}
	\label{dEh}
\end{eqnarray}
\begin{eqnarray}
	\delta E^{(f)}_{\mu \nu} &=& \frac{1}{4\pi}  \left\{ \mathcal{L}_F g^{\alpha \beta}(f_{\alpha \mu}F_{\beta \nu}+ F_{\alpha \mu}f_{\beta \nu})  \right. \nonumber \\
	&& \left. -\left( \mathcal{L}_F g_{\mu \nu}  - \mathcal{L}_{FF} g^{\alpha \beta} F_{\alpha \mu} F_{\beta \nu}   -  \mathcal{L}_{F_*F} F_* g_{\mu \nu} \right) \delta F^{(f)}  \right. \nonumber \\
&& \left. +\left(  \mathcal{L}_{FF_*} g^{\alpha \beta} F_{\alpha \mu} F_{\beta \nu}   +  \mathcal{L}_{F_*F_*} F_* g_{\mu \nu} \right) \delta F_*^{(f)}  \right\}.
	\label{dEf}
\end{eqnarray}
Here $\mathcal{L}_F =\partial \mathcal{L}/\partial F$, $\mathcal{L}_{FF} =\partial^2\mathcal{L}/\partial F^2$, and so on.
To derive the equations above, we use the fact that to first order
\begin{eqnarray}
	\tilde{F}=F+\delta F \nonumber \\
	\tilde{F_*}=F_*+\delta F_*
	\label{tF}
\end{eqnarray}
where $\delta F= \delta F^{(h)} +\delta F^{(f)}$ and $\delta F_*= \delta F_*^{(h)} +\delta F_*^{(f)}$, in which
\begin{equation}
	\delta F^{(h)} = -\frac{1}{2}g^{\alpha \beta}F_{\gamma \alpha}F_{\lambda \beta} h^{\gamma \lambda}
	\label{}
\end{equation}
\begin{equation}
	\delta F^{(f)} = \frac{1}{2}f_{\alpha \beta}F^{\alpha \beta}
	\label{}
\end{equation}
\begin{equation}
	\delta F_*^{(h)} = \frac{1}{2} F_* g_{\alpha \beta}h^{\alpha \beta}-g^{\alpha \beta}F_{*\gamma \alpha} F_{\lambda \beta} h^{\gamma \lambda}
	\label{}
\end{equation}
\begin{equation}
	\delta F_*^{(f)} = \frac{1}{2}f_{\alpha \beta}F_*^{\alpha \beta} .
	\label{}
\end{equation}
In the above, we also use the first order Taylor expansions
\begin{equation}
	\tilde{\mathcal{L}} = \mathcal{L} +\mathcal{L}_{F} \delta F + \mathcal{L}_{F_*} \delta F_*
	\label{}
\end{equation}
\begin{equation}
	\tilde{\mathcal{L}}_{\tilde{F}} = \mathcal{L}_F +\mathcal{L}_{FF} \delta F +\mathcal{L}_{FF_*} \delta F_* 
	\label{}
\end{equation}
\begin{equation}
	\tilde{\mathcal{L}}_{\tilde{F}_*} = \mathcal{L}_{F_*} +\mathcal{L}_{F_*F_*} \delta F_* + \mathcal{L}_{F_*F} \delta F .
	\label{}
\end{equation}
The perturbed Einstein-NLED equations (\ref{Einstein-h}) and (\ref{EM-f}) reduce to the Reissner-Nordst\"om results in \cite{Zerilli} when we choose $\mathcal{L}=F$ and $\Lambda=0$.


In addition to the above perturbed field equations, some of the background field equations are useful for this work.  A combination of the line element (\ref{LineElement}) and the Einstein equation in the form
\begin{equation}
	G_\mu{}^\nu=8\pi E_\mu{}^\nu,
	\label{EinsteinEqRaised}
\end{equation}
lead to the background field equations
\begin{equation}
	G_t{}^t=G_r{}^r= e^\nu \left( \frac{\nu'}{r} +\frac{1}{r^2} \right) -\frac{1}{r^2} =-2\mathcal{L}-\Lambda-\frac{2Q^2 \mathcal{L}_F}{r^4}
	\label{EERaised1}
\end{equation}
and
\begin{equation}
	G_\theta{}^\theta=G_\phi{}^\phi= \frac{e^\nu}{2} \left( \nu'' +\nu'^2 +\frac{2\nu'}{r} \right) =-2\mathcal{L}-\Lambda.
	\label{EERaised2}
\end{equation}
The above equations will be used extensively in the following sections.  For more details on the background field equations, see \cite{Nomura}.

An electrically charged black hole in NLED should satisfy some reasonable energy condition.  We denote $E_t{}^t =-\rho$, $E_r{}^r =p_r$,  $E_\theta{}^\theta =p_\theta$, and $E_\phi{}^\phi =p_\phi$, where $\rho$ is the energy density and $p_i$ ($i=r, ~\theta,~ \phi$) represents the pressure in the $i$ direction.  From Eq.\ (\ref{Emn}), we obtain
\begin{equation}
	\rho = - p_r=\frac{1}{4\pi}\left( \mathcal{L} + \frac{\Lambda}{2} + \frac{Q^2 \mathcal{L}_F}{ r^4} \right) 
	\label{}
\end{equation}
\begin{equation}
	p_\theta = p_\phi = -\frac{1}{4\pi}\left( \mathcal{L} + \frac{\Lambda}{2}  \right) .
	\label{}
\end{equation}
Using the above expressions for the energy density and pressure, we examine the following well-known options for the energy condition:

(a) The weak energy condition, where $\rho \ge 0$ and $\rho + p_i  \ge 0$, gives
\begin{equation}
	 \mathcal{L} + \frac{\Lambda}{2}  + \frac{Q^2 \mathcal{L}_F}{ r^4} \ge 0  ~~\text{and}~~   \mathcal{L}_F \ge 0.
	\label{}
\end{equation}

(b) The null energy condition, where $\rho + p_i \ge 0$, gives
\begin{equation}
	 \mathcal{L}_F \ge 0.
	\label{}
\end{equation}

(c) The dominant energy condition, where $\rho \ge  |p_i| $, gives

\hspace{0.7cm} $\bullet$ If $\mathcal{L} + \Lambda/2 \ge 0$, then  
\begin{equation}
	\mathcal{L}_F \ge 0. 
	\label{}
\end{equation}

\hspace{0.7cm} $\bullet$ If $\mathcal{L} + \Lambda/2 < 0$, then
\begin{equation}
	2 \mathcal{L} + \Lambda  + \frac{Q^2 \mathcal{L}_F}{ r^4} \ge 0 . 
	\label{}
\end{equation}

(d) The strong energy condition, where $\rho + p_i \ge 0$ and $\rho + \sum_{i}  p_i \ge 0 $, gives
\begin{equation}
	\mathcal{L}_F \ge 0    ~~\text{and}~~  -( 2 \mathcal{L}+ \Lambda)  \ge 0    .
	\label{}
\end{equation}
Note that all the above conditions force $\mathcal{L}_F \ge 0$.  Also, for the background field strength (\ref{Bianchi}) to be finite, we need $\mathcal{L}_F \ne 0$. Therefore, we will assume $\mathcal{L}_F > 0$ in the rest of the paper.

\section{Odd Parity Perturbations}
\label{Sec:opp}

The next step is to expand the perturbations $h_{\mu \nu}$ and $f_{\mu \nu}$ in tensor harmonics.  The odd parity (magnetic or axial) tensor expansion of the geometric perturbation is
\begin{equation}
 ||h_{\mu \nu}||=
\left[ {\begin{array}{cccc}
		0 & 0 & -h_0 \frac{1}{\sin \theta} \partial_\phi Y_{lm} & h_0 \sin \theta \partial_\theta Y_{lm}\\
		\\
		0 & 0 & -h_1 \frac{1}{\sin \theta} \partial_\phi Y_{lm} & h_1 \sin \theta \partial_\theta Y_{lm} \\ 
		\\
		sym & sym & h_2 \frac{1}{2\sin \theta} X_{lm} & -h_2 \frac{1}{2} \sin \theta W_{lm} \\ 
		\\
		sym & sym & sym & -h_2 \frac{1}{2} \sin \theta X_{lm}\\
\end{array} } \right],
\label{hmn}
\end{equation}
where $h_0$, $h_1$, and $h_2$ are functions of the time and radial coordinates only.  $Y_{lm}(\theta, \phi)$ are the spherical harmonics and 
\begin{eqnarray}
	X_{lm} &=& 2\partial_\phi \left(\partial_\phi -\cot \theta \right) Y_{lm}  \nonumber \\
	W_{lm} &=& \left( \partial^2_\theta -\cot \theta \partial_\theta - \frac{1}{\sin \theta} \partial^2_\phi \right) Y_{lm}. 
	\label{}
\end{eqnarray}
The integer $l \ge 2$ is the multipole number and $m=-l, \dots,0,\dots,l$. The freedom to make infinitesimal coordinate transformations allows us to fix the gauge in a way that $h_2=0$ (Regge-Weeler gauge\cite{RW}).

The odd parity tensor expansion of the NLED perturbation is
\begin{equation}
	||f_{\mu \nu}||=
	\left[ {\begin{array}{cccc}
			0 & 0 & \bar{f}_{02} \frac{1}{\sin \theta} \partial_\phi Y_{lm} & -\bar{f}_{02} \sin \theta \partial_\theta Y_{lm}\\
			\\
			0 & 0 &\bar{f}_{12} \frac{1}{\sin \theta} \partial_\phi Y_{lm} & -\bar{f}_{12} \sin \theta \partial_\theta Y_{lm} \\ 
			\\
			* & * & 0 & \bar{f}_{23} \sin \theta Y_{lm} \\ 
			\\
			* & * & * & 0\\
	\end{array} } \right],
\label{fmn-tensor}
\end{equation}
where $\bar{f}_{\mu \nu}$ denote angle-independent parts of $f_{\mu \nu}$.  The asterisk denotes the anti-symmetric components of the matrix.  

As noted by Zerilli in \cite{Zerilli}, the odd (even) parity geometric perturbations couple only to odd (even) parity electromagnetic perturbations. More specifically, when we combine odd with even parity, the Einstein-Maxwell equations lead to $\bar{f}_{\mu \nu}=0$.  We find this to be true for the NLED case considered here.  This, however, is not always true.  In a black hole with a magnetic monopole charge, odd parity geometric perturbations couple only to even parity electromagnetic perturbations and vice versa\cite{Nomura}.

Since the electromagnetic field tensor $\tilde{F}_{\mu \nu}$ is derived from a vector potential $\tilde{A}_\mu$, where  $\tilde{F}_{\mu \nu}=\tilde{A}_{\nu,\mu}- \tilde{A}_{\mu,\nu}$, we can write
\begin{equation}
    f_{\mu \nu}=a_{\nu,\mu}- a_{\mu,\nu},
	\label{fmn}
\end{equation}
where $a_\mu$ is the perturbed vector potential. This is equivalent to having the field equations of the form $f_{\mu\nu,\lambda}+f_{\lambda \mu,\nu}+f_{\nu\lambda, \mu}=0$.  These field equations lead to the following relations
\begin{eqnarray}
	\bar{f}_{12}=\frac{1}{l(l+1)} \partial_r  \bar{f}_{23}\nonumber \\
	\bar{f}_{02}=\frac{1}{l(l+1)} \partial_t  \bar{f}_{23}.
	\label{f23}
\end{eqnarray}

After inserting tensors (\ref{hmn}) and (\ref{fmn-tensor}) into Eq.\ (\ref{Einstein-h}), we obtain three equations from the components $r\theta$, $t\theta$, and $\theta \theta$ respectively
\begin{eqnarray}
	e^{-\nu} \partial_t^2 h_1 - e^{-\nu} \partial_r \partial_t h_0 +\frac{2}{r}e^{-\nu} \partial_t h_0 
	+ e^\nu \left( \nu'' +\nu'^2 + \frac{2\nu'}{r}\right)h_1 +2\lambda r^{-2}h_1 \nonumber \\
	=-4 \mathcal{L} h_1   -2\Lambda h_1 -4r^{-2}Q e^{-\nu} \mathcal{L}_F \bar{f}_{02} 
	\label{grav1}
\end{eqnarray}
\begin{eqnarray}
	-e^{\nu} \partial_r^2 h_0 + e^{\nu} \partial_r \partial_t h_1 +\frac{2}{r}e^{\nu} \partial_t h_1 
	+ e^\nu \left( \nu'' +\nu'^2 + \frac{2\nu'}{r} +\frac{2}{r^2}\right)h_0 +2\lambda r^{-2}h_0 \nonumber \\
	=-4 \mathcal{L} h_0  -2\Lambda h_0 -4r^{-2}Q e^\nu \mathcal{L}_F \bar{f}_{12} 
	\label{grav2}
\end{eqnarray}
\begin{equation}
	-e^{-\nu}\partial_t h_0  +e^\nu \partial_r h_1 + e^\nu \nu' h_1 = 0,
	\label{grav3}
\end{equation}
where $\lambda= \frac{1}{2}[l(l+1)-2]$. Throughout this paper, we use prime to denote the derivative with respect to the radial coordinate $r$. In addition, from the $rr$ component of the perturbed Einstein equation, we find that 
\begin{equation}
	\mathcal{L}_{FF_*}=0
	\label{grav4}
\end{equation}
when $F_*=0$, which is the case for an electric charge.
This constraint on $\mathcal{L}$ is also noticed by the authors of \cite{Nomura}, where they suggest a general form for $\mathcal{L}$ in which
\begin{equation}
	\mathcal{L}(F, F_*)=\mathcal{L}_0(F)+ \sum_{n=2}^\infty \frac{1}{n!} \mathcal{L}_n(F) F_*^n.
	\label{}
\end{equation}
Inserting tensors (\ref{hmn}) and (\ref{fmn-tensor}) into Eq.\ (\ref{EM-f}) and using Eq.\ (\ref{f23}), we obtain
\begin{eqnarray}
	\mathcal{L}_F e^\nu \partial_r \left(e^\nu \partial_r \bar{f}_{23}\right) - \mathcal{L}_F \partial_t^2 \bar{f}_{23}  + \mathcal{L}_F' e^{2\nu} \partial_r \bar{f}_{23} - \frac{l(l+1)}{r^6}e^\nu \left( r^4 \mathcal{L}_F +Q^2 \mathcal{L}_{F_*F_*} \right) \bar{f}_{23} \nonumber \\
	= \frac{l(l+1)}{r^2}  e^\nu Q \mathcal{L}_F \left[-\partial_t h_1+ r^2\partial_r (h_0/r^2) \right] +\frac{l(l+1)}{r^2}  e^\nu \left(Q' \mathcal{L}_F +Q \mathcal{L}_F'  \right) h_0.
	\label{em2}
\end{eqnarray}
We then solve Eq.\ (\ref{grav3}) for $h_0$ and substitute it into Eq.\ (\ref{grav1}).  After defining $R_{lm}^{(odd)} = (1/r)e^\nu h_1$, $f_{lm}^{(odd)} = \sqrt{\mathcal{L}_F}\bar{f}_{23}/[l(l+1)]$, using the tortoise coordinate $r_*$ where $dr_*/dr=e^{-\nu}$, and using Eq.\ (\ref{EERaised2}), we get
\begin{equation}
	\frac{d^2 R_{lm}^{(odd)}}{dr_*^2} +  
	\left\{ \omega^2 - e^\nu \left[ \frac{2\lambda}{r^2} +e^\nu \left(- \frac{\nu'}{r} + \frac{2}{r^2}\right)\right] \right\} R_{lm}^{(odd)} =-\frac{4i\omega}{r^3}Q e^\nu \sqrt{\mathcal{L}_F}  f_{lm}^{(odd)}   
	\label{SEodd1}
\end{equation}   
while Eq.\ (\ref{em2}) becomes
\begin{flushleft}
	\begin{eqnarray}
		&&\frac{d^2 f_{lm}^{(odd)}}{dr_*^2}+ \left\{\omega^2  - e^\nu \left[ \frac{l(l+1)}{r^6\mathcal{L}_F} \left( r^4 \mathcal{L}_F + Q^2 \mathcal{L}_{F_*F_*} \right) \right. \right. \nonumber \\
		&&~~~~~~~~~~~~~~~~~~~~~~~~~~\left. \left.  +\frac{4}{r^4}Q^2\mathcal{L}_F   - \frac{e^{-\nu}}{4 \mathcal{L}_F^2}\left(\left(\frac{d \mathcal{L}_F}{dr_*}\right)^2-2 \mathcal{L}_F \frac{d^2 \mathcal{L}_F}{dr_*^2} \right)   \right] \right\} f_{lm}^{(odd)}
		\nonumber \\ 
		&&= -\frac{2\lambda}{i \omega r^3} Q e^\nu \sqrt{\mathcal{L}_F}   R_{lm}^{(odd)}. 
		\label{SEodd2}
	\end{eqnarray}     
\end{flushleft}
In the above two equations, we assume all field functions depend on time as $e^{-i\omega t}$, where $\omega$ is the quasinormal mode frequency of the perturbations.  This is formally equivalent to a Fourier transform of the field functions where $\partial_t \rightarrow -i \omega$. Recall, we are also requiring $\mathcal{L}_F  > 0$, which makes $\sqrt{\mathcal{L}_F}$ well-defined.  Equations (\ref{SEodd1}) and (\ref{SEodd2}) reduce to the Reissner-Nordstr\"om wave equations when $\mathcal{L}=F$, $\Lambda=0$, and $e^\nu=1-\frac{2M}{r}+\frac{q^2}{r^2}$.


Wave equations (\ref{SEodd1}) and (\ref{SEodd2}) are valid for multipole numbers of $l \ge 2$. In the case of $l=1$, where $\lambda =0$, wave equation (\ref{SEodd2}) decouples from (\ref{SEodd1}).  In this case, only the  electromagnetic perturbations are dynamical degrees of freedom and the perturbations are completely described by Eq.\ (\ref{SEodd2}).  This is because $h_0$ is only defined for  $l \ge 2$.  As a result, for $l=1$, $h_1$ (and consequently  $R_{lm}^{(odd)}$) is no longer a physical degree of freedom.  This can be shown by simplifying Eq.\ (\ref{grav1}) using the background field equation (\ref{EERaised2}) and taking $h_0$ and $\lambda$ to be zero, which gives
\begin{equation}
	h_1=\frac{2 q}{i \omega r^2} \bar{f}_{23}= \frac{4 q}{i \omega r^2 \sqrt{\mathcal{L}_F}} f_{lm}^{(odd)}.
	\label{}
\end{equation}

\section{Stability for Odd Parity Perturbations}
\label{Sec:SOdd}

To derive the stability condition for odd parity perturbations, we follow the method in \cite{Moreno}.  Defining $\hat{f}_{lm}^{(odd)}= \frac{2i\omega}{\sqrt{2\lambda}}f_{lm}^{(odd)}$, we can rewrite the wave equations (\ref{SEodd1}) and (\ref{SEodd2}) as
\begin{equation}
	e^{-\nu} \left[ r \frac{d}{dr_*}\left( \frac{1}{r^2}\frac{d }{dr_*}\left( r R_{lm}^{(odd)}\right) \right) + \omega^2 R_{lm}^{(odd)}\right]- V{}_{11}^\text{I} R_{lm}^{(odd)}- V{}_{12}^\text{I} \hat{f}_{lm}^{(odd)}=0 
	\label{Moreno1}
\end{equation}   
\begin{equation}
	e^{-\nu} \left[ \frac{1}{\sqrt{\mathcal{L}_F}} \frac{d}{dr_*}\left( \mathcal{L}_F \frac{d }{dr_*}\left(\frac{  \hat{f}_{lm}^{(odd)}}{\sqrt{\mathcal{L}_F}}\right) \right) + \omega^2 \hat{f}_{lm}^{(odd)}\right]- V{}_{22}^\text{I} \hat{f}_{lm}^{(odd)}- V{}_{21}^\text{I} R_{lm}^{(odd)}=0 ,
	\label{Moreno2}
\end{equation}   
where
\begin{equation}
    V{}_{11}^\text{I} = \frac{2\lambda}{r^2}
	\label{Moreno3}
\end{equation}   
\begin{equation}
	V{}_{12}^\text{I} = V{}_{21}^\text{I}= -\frac{2\sqrt{2\lambda \mathcal{L}_F}Q}{ r^3} 
	\label{Moreno4}
\end{equation} 
\begin{equation}
	V{}_{22}^\text{I} =  \frac{l(l+1)}{r^6\mathcal{L}_F} \left( r^4 \mathcal{L}_F + Q^2 \mathcal{L}_{F_*F_*} \right)  +\frac{4}{r^4}Q^2\mathcal{L}_F .
	\label{Moreno5}
\end{equation} 
Equations (\ref{Moreno1}-\ref{Moreno5}) are in good agreement with the results found in \cite{Moreno}.  The contribution from including the Hodge dual scalar invariant $F_*$ can be found in the potential (\ref{Moreno5}).

The stability condition in \cite{Moreno}, given by requiring  the potential matrix 
\begin{equation}
	V^\text{I}=
	\left[ {\begin{array}{cc}
			V{}_{11}^\text{I} & V{}_{12}^\text{I} \\
			V{}_{21}^\text{I} & V{}_{22}^\text{I}  \\
	\end{array} } \right]
	\label{}
\end{equation}
to be positive-definite, will be modified due to the inclusion of Hodge dual fields.  We require the  determinant and trace 
\begin{equation}
	\text{det}(V^\text{I}) =    \frac{4 \lambda (\lambda+1)}{r^8\mathcal{L}_F} \left( r^4 \mathcal{L}_F + Q^2 \mathcal{L}_{F_*F_*} \right)  
	\label{}
\end{equation} 
\begin{equation}
	\text{tr}(V^\text{I}) =   \frac{2\lambda}{r^2} +  \frac{2(\lambda+1)}{r^6\mathcal{L}_F} \left( r^4 \mathcal{L}_F + Q^2 \mathcal{L}_{F_*F_*} \right)  +\frac{4}{r^4}Q^2\mathcal{L}_F 
	\label{}
\end{equation} 
be positive.  Since $\lambda >0$ and $\mathcal{L}_F >0$, this gives the stability condition as
\begin{equation}
	 1 + \frac{Q^2 \mathcal{L}_{F_*F_*} }{r^4 \mathcal{L}_F} =  1 -2 F \frac{\mathcal{L}_{F_*F_*} }{\mathcal{L}_F}  > 0.
	\label{stability-odd-0}
\end{equation} 

For $l=1$, where $\lambda=0$, the perturbations are completely described by Eq.\ (\ref{Moreno2}) where $V{}_{21}^\text{I}=0$. Therefore, for the black hole stability against electromagnetic perturbations with $l=1$, the only requirement is  $V{}_{22}^\text{I} >0$.  This gives us a condition that holds when Eq.\ (\ref{stability-odd-0}) is satisfied.  Therefore, the stability condition (\ref{stability-odd-0}) can be applied to multipole numbers $l\ge 1$.

\section{Even Parity Perturbations}
\label{Sec:epp}

The even parity (electric or polar) tensor expansion of the geometric perturbation is
\begin{equation}
	||h_{\mu \nu}||=
	\left[ {\begin{array}{cccc}
			e^\nu H_0 Y_{lm} & H_1 Y_{lm}  & h_0^{(e)} \partial_\theta Y_{lm} & h_0^{(e)} \partial_\phi Y_{lm} \\
			\\
			sym & e^{-\nu} H_2 Y_{lm} & h_1^{(e)} \partial_\theta Y_{lm} & h_1^{(e)} \partial_\phi Y_{lm}  \\ 
			\\
			sym & sym & r^2\left( K Y_{lm} +G \partial_\theta Y_{lm} \right) & \frac{1}{2} r^2 G X_{lm} \\ 
			\\
			sym & sym & sym &  r^2 \sin^2\theta \left[ K Y_{lm} +G \left( \partial_\theta^2 Y_{lm} -W_{lm} \right) \right] \\
	\end{array} } \right],
	\label{hmn-even}
\end{equation}
where $H_0$, $H_1$, $h_0^{(e)}$, $h_1^{(e)}$, $K$, and $G$ are functions of the time and radial coordinates only.  In the Regge-Wheeler gauge, $h_0^{(e)}\equiv h_1^{(e)}\equiv G \equiv 0$.  The even parity tensor expansion of the NLED perturbation is
\begin{equation}
	||f_{\mu \nu}||=
	\left[ {\begin{array}{cccc}
			0 & \bar{f}_{01}  Y_{lm}  & \bar{f}_{02} \partial_\theta Y_{lm} & \bar{f}_{02} \partial_\phi Y_{lm}\\
			\\
			* & 0 &\bar{f}_{12}  \partial_\theta Y_{lm} & \bar{f}_{12}  \partial_\phi Y_{lm} \\ 
			\\
			* & * & 0 & 0 \\ 
			\\
			* & * & 0 & 0\\
	\end{array} } \right],
	\label{fmn-even}
\end{equation}
where $\bar{f}_{01}$, $\bar{f}_{02}$, and $\bar{f}_{12}$ are functions of the time and radial coordinates only.  
One can use the same idea as in Eq.\ (\ref{fmn}) to find the homogeneous Maxwell equation
\begin{equation}
	\bar{f}_{01}   = \partial_r \bar{f}_{02}-\partial_t \bar{f}_{12} 
	\label{homogen-even}
\end{equation}
for even parity perturbations.

The $tt$, $rr$, a combination of $ \theta \theta $ and $ \phi \phi $\footnote{The $ \theta \theta $ and $ \phi \phi $ components of Eq.\ (\ref{Einstein-h}) are the same with the exception of one angle-dependent term in each component, both of which involve $H_2-H_0$.  However,  these angle-dependent terms can be combined to become angle-independent by simply taking the average of the $\theta \theta $ and $ \phi \phi $  components.}, $t r$, $t \theta $, $r \theta $, and $ \theta \phi $ components of Eq.\ (\ref{Einstein-h}) are respectively
\begin{eqnarray}
	&&e^{2\nu}\left[2 \partial_r^2 K -\frac{2}{r} \partial_r H_2 +\left( \nu'+\frac{6}{r}\right) \partial_rK  -2\left( \frac{1}{r^2} +\frac{\nu'}{r} \right) (H_0+H_2) \right] 	\nonumber \\ 
	&&~~+e^\nu \left(\frac{2}{r^2} H_0  -\frac{2\lambda}{r^2} K -\frac{l(l+1)}{r^2} H_2 \right) =\frac{2Q}{r^8}e^\nu (Q H_2  +2r^2 \bar{f}_{01}) \left( r^4 \mathcal{L}_F-Q^2 \mathcal{L}_{FF}  \right)
	\nonumber \\
	&&~~~~~~~~~~~~~~~~~~~~~~~~~~~~~~~~~~~~~~~~~~~~~~~~+2 e^\nu H_0 \left( \Lambda +2 \mathcal{L}+ \frac{Q^2}{r^4} \mathcal{L}_{F} + \frac{Q^4}{r^8} \mathcal{L}_{FF} \right)
	\label{gravEven00}
\end{eqnarray}
\begin{eqnarray}
	&&2 e^{-2\nu} \partial_t^2 K -\frac{4}{r} e^{-\nu} \partial_t H_1 +\frac{2}{r} \partial_r H_0 -\left( \nu' +\frac{2}{r} \right)\partial_r K +\frac{2}{r^2} e^{-\nu}  H_2 -\frac{l(l+1)}{r^2} e^{-\nu} H_0 +\frac{2\lambda}{r^2} e^{-\nu} K  \nonumber \\
	&&    =\frac{2Q}{r^8}e^{-\nu} (Q H_0  -2r^2 \bar{f}_{01}) \left( r^4 \mathcal{L}_F-Q^2 \mathcal{L}_{FF}  \right) +2 e^{-\nu} H_2 \left( \Lambda +2 \mathcal{L}+ \frac{Q^2}{r^4} \mathcal{L}_{F} + \frac{Q^4}{r^8} \mathcal{L}_{FF} \right)
	\label{gravEven11}
\end{eqnarray}
\begin{eqnarray}
	&&r^2 \left[ e^{-\nu} \partial_t^2 K -e^{\nu}\partial_r^2 K -e^\nu \left( \nu' +\frac{2}{r} \right)\partial_r K- \left( \nu' +\frac{2}{r} \right)\partial_t H_1+ e^{-\nu}  \partial_t^2 H_2 -2\partial_r \partial_t H_1 + e^{\nu}  \partial_r^2 H_0 \right. \nonumber \\
	&& \left. +e^\nu \left( \frac{1}{2}\nu'+\frac{1}{r} \right)\partial_r H_2 +e^\nu \left( \frac{3}{2}\nu'+\frac{1}{r} \right)\partial_r H_0 +\frac{l(l+1)}{2r^2}  (H_2 -H_0) \right.  \nonumber \\
	&&\left. +e^\nu \left(\nu''+\nu'^2+\frac{2\nu'}{r}  \right)  (H_2-K) \right] = 2r^2(\Lambda+2\mathcal{L})K 
	+4 Q \mathcal{L}_F  \bar{f}_{01} +2\frac{Q^2}{r^2}\mathcal{L}_F(H_2-H_0)
	\label{gravEven22+33/2}
\end{eqnarray}
\begin{eqnarray}
	&&2\partial_r \partial_t K-\frac{2}{r} \partial_t H_2 - \left( \nu' -\frac{2}{r} \right) \partial_t K-e^\nu \left( \frac{2\nu'}{r} + \frac{2}{r^2} \right) H_1 - \frac{2\lambda}{r^2} H_1 \nonumber \\
	&&= 2\left( \Lambda+2\mathcal{L}+\frac{2Q^2}{r^4}\mathcal{L}_F \right) H_1
	\label{gravEven10}
\end{eqnarray}
\begin{equation}
	-e^\nu \partial_r H_1 + \partial_t K + \partial_t H_2 - e^\nu \nu' H_1 =-\frac{4Q}{r^2} e^\nu  \mathcal{L}_F \bar{f}_{12}
	\label{gravEven20}
\end{equation}
\begin{equation}
	e^{-\nu} \partial_t H_1 - \partial_r H_0 + \partial_r K - \left( \frac{\nu'}{2} +\frac{1}{r}\right) H_2 - \left( \frac{\nu'}{2} -\frac{1}{r}\right) H_0 =-\frac{4Q}{r^2} e^{-\nu}  \mathcal{L}_F \bar{f}_{02}
	\label{gravEven12}
\end{equation}
\begin{equation}
    H_2-H_0=0.
	\label{gravEven23}
\end{equation}
Note that Eqs.\ (\ref{gravEven00}), (\ref{gravEven11}), and (\ref{gravEven22+33/2}) do not reduce to Zerilli's results in \cite{Zerilli} for the Reissner-Nordstr\"om case. For an explanation, see Appendix C. 

The $r$, $\theta$ or $\phi$, and $t$ components of the perturbed NLED equation (\ref{EM-f}) are respectively
\begin{eqnarray}
	&&\left(\mathcal{L}_F -\frac{Q^2}{r^4} \mathcal{L}_{FF} \right) \partial_t \bar{f}_{01} -\frac{1}{r^2} l(l+1) e^\nu \mathcal{L}_F \bar{f}_{12}  \nonumber \\
	&&=\frac{Q}{2r^2} \left[ \left(\mathcal{L}_F-\frac{Q^2}{r^4} \mathcal{L}_{FF}\right)\left(\partial_t H_0-\partial_t H_2\right)+2 \mathcal{L}_F \partial_t K  \right]
	\label{EMEven1}
\end{eqnarray}
\begin{equation}
	e^{-\nu} \mathcal{L}_F \partial_t \bar{f}_{02} -  \partial_r \left(e^\nu  \mathcal{L}_F \bar{f}_{12}  \right) =0
	\label{EMEven2}
\end{equation}
\begin{eqnarray}
	&&\frac{1}{r^2} \partial_r \left[\left(r^2 \mathcal{L}_F -\frac{Q^2}{r^2} \mathcal{L}_{FF}\right) \bar{f}_{01}\right] -\frac{1}{r^2} l(l+1) e^{-\nu} \mathcal{L}_F \bar{f}_{02}\nonumber \\
	&&=\frac{1}{2r^2} \partial_r \left[ Q \left(1-2  \mathcal{L}_{F}+\frac{Q^2}{r^4} \mathcal{L}_{FF}\right)\left( H_2- H_0\right)+2 Q K  \right].
	\label{EMEven3}
\end{eqnarray}
Note that, in addition to Eq.\ (\ref{EMEven2}), the  $\theta$ or $\phi$ component of the perturbed NLED equation requires
\begin{equation}
	\left(\mathcal{L}_{F_*} -\frac{Q^2}{r^4} \mathcal{L}_{F_* F} \right) \bar{f}_{01} + \mathcal{L}_{F_*}\left( \partial_t \bar{f}_{12}  - \partial_r \bar{f}_{02}\right) -\mathcal{L}_{F_*}'\bar{f}_{02} 
	=\frac{Q^3}{2r^6} \mathcal{L}_{F_* F} (H_2-H_0),
	\label{}
\end{equation}
which is satisfied only if $\mathcal{L}_{F_* F}=0$ and $\bar{f}_{01}+\partial_t \bar{f}_{12}  - \partial_r \bar{f}_{02}=0$ that we already determined in  Eqs.\  (\ref{grav4}) and (\ref{homogen-even}).  This provides a good consistency check.

Note that the Hodge dual of the electromagnetic field does not appear anywhere in the equations (\ref{gravEven00}-\ref{EMEven3}).  Therefore, the even parity perturbations are unaltered by the inclusion of Hodge dual fields.  So, equations (\ref{gravEven00}-\ref{EMEven3}) should, and do, reduce to a pair of coupled Schr\"odinger-type wave equations, which agree with those in \cite{Moreno}.  Likewise, the stability conditions for even parity perturbations do not change from those that appear in \cite{Moreno}. 

Since our method is different than that used in \cite{Moreno}, we include the derivation of the wave equations in Appendices A and B.  In Appendix A, we use a method similar to that in \cite{Zerilli} to find the wave equations that  reduce to those in \cite{Zerilli} in the Reissner-N\"ordstrom case.  In Appendix B, we show how to rewrite the wave equations in the form in which they appear in \cite{Moreno} and are more suitable for stability analysis.

\section{An Application: Born-Infeld theory}
\label{Sec:application}

In this section, we provide an example of a viable theory that involves Hodge dual fields.   
In the original work of Born and Infeld \cite{Born}, they removed the divergence of an electron's self-energy in classical electrodynamics by introducing a nonlinear Lagrangian density of the form
\begin{equation}
	\mathcal{L}(F, F_*) =\mu^4 \sqrt{1+\frac{2F}{\mu^4}-\frac{F_*^2}{\mu^8}}-\mu^4,
	\label{Born-L}
\end{equation}
where $\mu$ is a scale parameter of dimension mass.  It is easy to see this Lagrangian density reduces to Maxwell's when $F/\mu^4 \ll 1$.  Born-Infeld theory in curved spacetime (Einstein-Born-Infeld gravity) has been explored in the literature.  For electrically charged black hole solutions, see for example \cite{BIbh1, BIbh2}.

\begin{figure}[th!]
	\begin{center}
		\includegraphics[height=7cm]{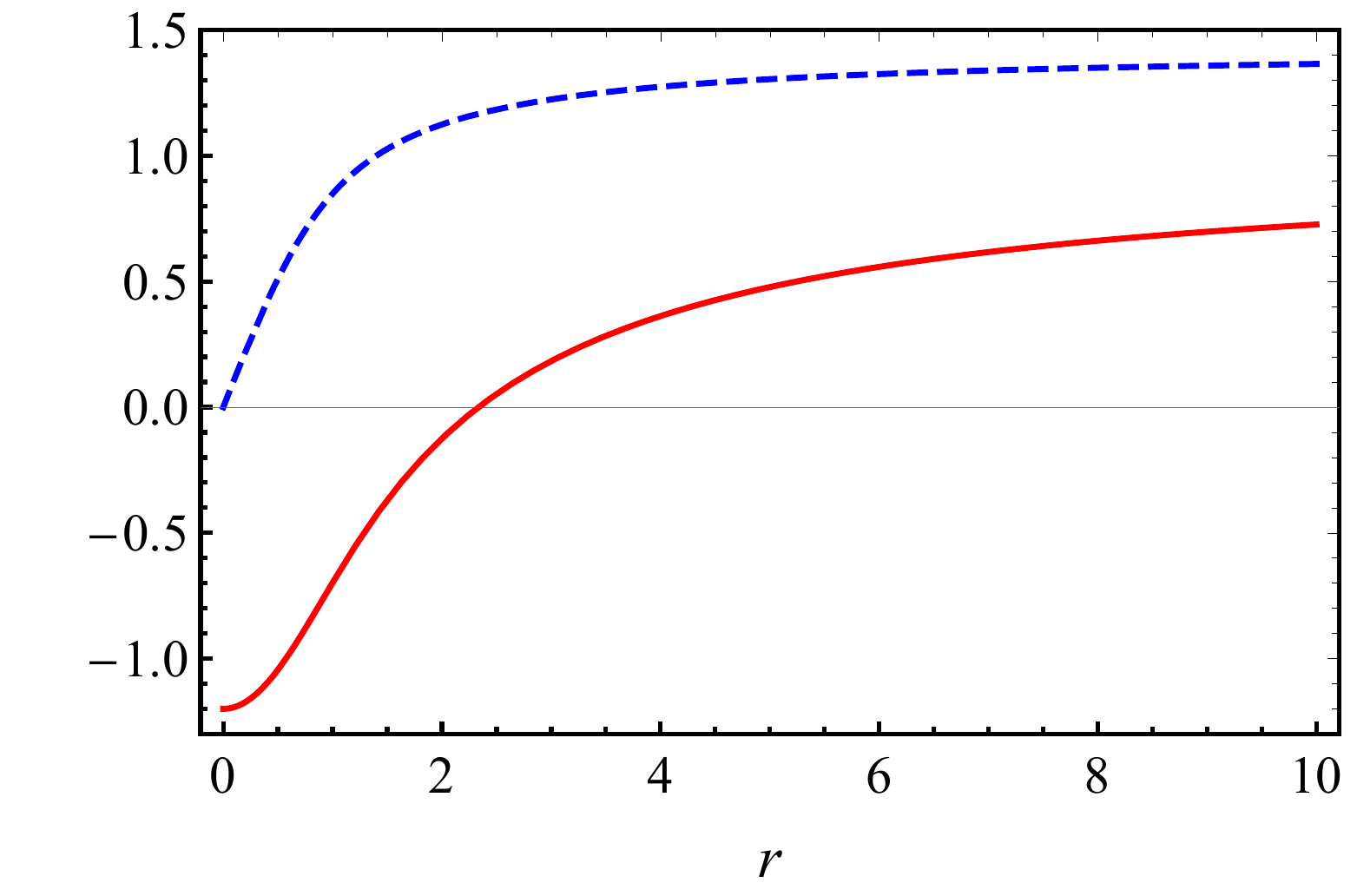}
	\end{center}
	\vspace{-0.7cm}
	\caption{\footnotesize The plot shows $M$, in dashed blue, and $e^\nu$, in solid red, as a function of the radial coordinate $r$.  Here $\mu=1$, $|q|=1.1/\mu^2$, and $\Lambda=0$.}
	\label{figM}
\end{figure}

If we use a metric function of the form
\begin{equation}
	e^\nu= 1-\frac{2 M(r)}{r}-\frac{\Lambda}{3}r^2,
	\label{}
\end{equation}
together with the background field equations (\ref{EERaised1}) and (\ref{EERaised2}), we find
\begin{equation}
	\frac{M'}{r^2}=\mathcal{L}-2 F \mathcal{L}_F  
	\label{M'eq}
\end{equation}
\begin{equation}
	\frac{M''}{r}=2 \mathcal{L}.
	\label{M''eq}
\end{equation}
We then take the derivative of Eq.\ (\ref{Born-L}) with respect to $F$. Replacing $F$ and $F_*$ with their background values of  $-\frac{q^2}{2r^4 \mathcal{L}_F^2}$ and $0$ gives an equation in $\mathcal{L}_F$.  Using $\mathcal{L}_F>0$ as required by the energy conditions listed at the end of Section \ref{Sec:WE}, we obtain
\begin{equation}
	\mathcal{L}_F (r)=\left(1+\frac{q^2}{\mu^4 r^4} \right) ^{1/2}.
	\label{LFr}
\end{equation}
We can use Eq.\ (\ref{LFr}) to  write $F$, and consequently $\mathcal{L}$, as functions of $r$ only. This allows us to integrate the background field equation (\ref{M'eq}) to get
 \begin{equation}
 	M(r)=-\frac{1}{3}\mu^4 r^3 +\frac{1}{3}\mu^2 r \sqrt{q^2 + \mu^4 r^4} -\frac{2}{3}  \sqrt{i \mu^2 q^3}~ \mathbb{F} \left[\arcsin \left(-i r\sqrt{\mu^2/q}\right) \bigg{|} -1\right],
 	\label{}
 \end{equation}
where $\mathbb{F}(\varphi |k^2)$ is the elliptic integral of the first kind. In the asymptotic region of $r \rightarrow 0$, $M(r) \approx \sqrt{q^2} \mu^2 r$.   As $r \rightarrow \infty$, $M(r)$ approaches a positive constant.  Therefore, for $|q|> \frac{1}{2\mu^2}$, the metric function $e^\nu$ starts with a finite negative value of $1- 2\sqrt{q^2}\mu^2$  at $r=0$ and approaches $1$ (for $\Lambda=0$) as $r \rightarrow \infty$. This provides us with the spacetime of a black hole.  We show the behavior of $M$ and $e^\nu$ as a function of $r$ in Figure \ref{figM}.  

For this black hole, the stability condition (\ref{stability-odd-0}) translates to
\begin{equation}
	\mu^4 r^4 >0.
	\label{c0}
\end{equation}
Since this is always true,  we can conclude that electrically charged black hole solutions in Einstein-Born-Infeld gravity are stable against odd parity perturbations.  This includes purely electromagnetic perturbations with $l=1$ as discussed at the end of Sec.\ \ref{Sec:SOdd}.

For even parity perturbations, we can use the same stability conditions derived in \cite{Moreno}.  These are
\begin{eqnarray}
	&&\mathcal{H}<0 \label{c1} \\
	&&\mathcal{H}_P > 0  \label{c2} \\
	&& 0 < e^\nu \left(1+2 \frac{\mathcal{H}_{PP}}{\mathcal{H}_P}P\right)\le 3,
	\label{c3}
\end{eqnarray}
where $\mathcal{H}=2 F \mathcal{L}_F-\mathcal{L}$ and $P=-\frac{q^2}{2r^4}$.  These conditions apply to the region outside the event horizon.  We combine Eqs.\ (\ref{ScalarF}), (\ref{Born-L}), and (\ref{LFr}) to get
\begin{equation}
	\mathcal{H}(P)=\mu^4\left(1- \sqrt{1-2\frac{P}{\mu^4}} \right).
	\label{Hx}
\end{equation}
The stability condition (\ref{c1}) gives 
\begin{equation}
	\left(1+\frac{q^2}{\mu^4 r^4} \right)^{1/2} > 1,
	\label{Hless0}
\end{equation}
which is the same as $\mathcal{L}_F >1$.   This is true as long as $q$ is not zero.  It is easy to show that the condition (\ref{c2}) is satisfied when  inequality (\ref{Hless0}) holds.  The condition (\ref{c3}) gives
\begin{equation}
	0 < e^\nu  	\left(1+\frac{q^2}{\mu^4 r^4} \right)^{-1} \le 3.
	\label{c4'}
\end{equation}
For $\Lambda=0$, since $0 < e^\nu < 1$ outside the event horizon, condition (\ref{c4'}) is always satisfied. We conclude that electrically charged black holes in an asymptotically Minkowski spacetime  in Einstein-Born-Infeld gravity are stable.

\section{Summary and Conclusion}
\label{Sec:conclusions}

We studied the perturbations of the Einstein equation coupled to general NLED for a spherically symmetric black hole solution with electric charge. We also included the cosmological constant and the Hodge dual of the electromagnetic field strength tensor in our calculations. The NLED Lagrangian density is a generic function of the Lorentz invariant scalar quantities of the electromagnetic fields, i.e. $F$ and $F_*$. The wave equations for odd and even parity perturbations of gravitational and NLED fields were derived.  For each parity, we reduced the Einstein-NLED field equations to two coupled Schr\"odinger-type equations, one of which determines the gravitational and the other the NLED field oscillations. 

Our results are consistent with those found in \cite{Moreno}, although we did not use the gauge-invariant technique utilized by Moreno and Sarbach in \cite{Moreno}.  Our method, where we fixed the gauge early on, is more in line with the work done by Nomura {\it et al.} in \cite{Nomura} and by Zerilli in \cite{Zerilli}.  We also included the Hodge dual of the electromagnetic field strength tensor, which was ignored in \cite{Moreno}.    In addition, all our equations reduce to the correct results for the Reissner-Nordstr\"om case when we use Maxwell's Lagrangian density ($\mathcal{L}=F$) and take the cosmological constant $\Lambda$ to be zero.  

The inclusion of the Hodge dual of the electromagnetic field modifies the results of \cite{Moreno} only for odd parity perturbations.  The even parity perturbations stay unaltered.  Therefore, we conclude that the inclusion of $F_*$ does not change the stability conditions for even parity perturbations that were explored earlier in the literature. We provided new stability conditions for the odd parity perturbations that include the Hodge dual of the electromagnetic field.

\vskip .5cm


\appendix
\section{Derivation of Even Parity Wave Equations}
\label{Sec:WEEven}

In this appendix, we show how to use equations (\ref{gravEven00}-\ref{EMEven3}) to derive two coupled Schr\"odinger-type wave equations for even parity perturbations.

First, we find $\bar{f}_{01}$ and $\bar{f}_{02}$ in terms of $\bar{f}_{12}$ by solving Eqs.\ (\ref{EMEven1}) and  (\ref{EMEven2}) respectively.  We then substitute these values to Eq.\ (\ref{homogen-even}) to find a second order differential equation for $\bar{f}_{12}$:  
\begin{eqnarray}
	&&\partial_{r}^2 (e^\nu \mathcal{L}_F \bar{f}_{12}) +\left( \nu' -\frac{ \mathcal{L}_F'}{ \mathcal{L}_F}  \right) \partial_{r} (e^\nu \mathcal{L}_F \bar{f}_{12})- e^{-2\nu}\partial_{t}^2  (e^\nu \mathcal{L}_F \bar{f}_{12})-  \frac{l(l+1) r^2 \mathcal{L}_F}{e^\nu \left(r^4 \mathcal{L}_F - Q^2 \mathcal{L}_{FF} \right)}      (e^\nu \mathcal{L}_F \bar{f}_{12})
	\nonumber \\ 
	&& ~~~~~~~~~~~~~~~~~~~~~~~~~~~~~~~~~~~~~~~~~~~~~~~~~~~~~~~ = \frac{r^2 Q \mathcal{L}_F^2}{e^\nu \left( r^4 \mathcal{L}_F - Q^2 \mathcal{L}_{FF} \right)} \partial_t K.
	\label{f12-eq}
\end{eqnarray}     
We define $f_{lm}^{(even)} = e^\nu \sqrt{\mathcal{L}_F}\bar{f}_{12}$, and use the tortoise coordinate $r_*$ where $dr_*/dr=e^{-\nu}$, to find
\begin{eqnarray}
	&&\partial_{r_*}^2 f_{lm}^{(even)} - \partial_{t}^2 f_{lm}^{(even)} 
	- e^\nu \left\{ \frac{l(l+1) r^2 \mathcal{L}_F}{\left( r^4 \mathcal{L}_F - Q^2 \mathcal{L}_{FF} \right)}    + \frac{e^{-\nu}}{4 \mathcal{L}_F^2}\left[3\left(\frac{d\mathcal{L}_F}{dr_*}\right)^2-2 \mathcal{L}_F \frac{d^2 \mathcal{L}_F}{dr_*^2}\right]   \right\} f_{lm}^{(even)}
	\nonumber \\ 
	&&\hspace{9cm} =  \frac{r^2 e^\nu Q \mathcal{L}_F^{3/2}}{\left(r^4 \mathcal{L}_F - Q^2 \mathcal{L}_{FF} \right)} \partial_t K.
	\label{}
\end{eqnarray}

In the remainder of this section, we assume all field functions depend on time as $e^{-i\omega t}$, where $\omega$ is a complex constant that turns out to be the quasinormal mode frequency of the perturbations.    We now look at the geometric perturbation equations (\ref{gravEven00}-\ref{gravEven23}).  We use Eq.\ (\ref{gravEven23}) to eliminate $H_2$ in Eqs.\ (\ref{gravEven10}-\ref{gravEven12}).  We then substitute $\partial_r K$ and $\partial_r H_1$, as given by these equations, into Eq.\ (\ref{gravEven11}).  This gives  an algebraic equation that involves $H_0$, $H_1$, $K$ and the electromagnetic functions $\bar{f}_{01}$, $\bar{f}_{02}$, $\bar{f}_{12}$.  We now solve this equation for $H_0$ and substitute into Eqs.\ (\ref{gravEven10}) and (\ref{gravEven20}).  Using Eqs.\ (\ref{EMEven1}) and (\ref{EMEven2}), we replace $\bar{f}_{01}$ and $\bar{f}_{02}$ with  $\bar{f}_{12}$.  This procedure gives the following two equations
\begin{equation}
	\frac{dK}{dr} = \alpha_\omega (r) K +\omega^{-1} \beta_\omega(r) H_1 +S_1
	\label{dK}
\end{equation}
\begin{equation}
	\omega^{-1} \frac{dH_1}{dr}= \gamma_\omega (r) K +\omega^{-1} \delta_\omega(r) H_1 +S_2,
	\label{dH1}
\end{equation}
where
\begin{eqnarray}
	\alpha_\omega (r)=\frac{4 e^\nu Q^2 \mathcal{L}_F  -r^2\xi(e^\nu-\lambda-1)-2r^2(\lambda+1)^2+2r^2 e^\nu(2\lambda+1) -2\omega^2 r^4 }{r^3 e^\nu \xi(r)} 
	\label{alpha}
\end{eqnarray}
\begin{equation}
	\beta_\omega (r)=2i\frac{(\lambda+1)\left[(\lambda+1)-e^\nu \right]+\omega^2 r^2}{r^2\xi(r)}
	\label{beta}
\end{equation}
\begin{equation}
	\gamma_\omega (r)=i\frac{-8 e^\nu  Q^2 \mathcal{L}_F+r^2(\xi-2\lambda-2)^2-4r^2 e^\nu (2\lambda+1) +4\omega^2 r^4  }{2 r^2 e^{2\nu}\xi(r)}
	\label{gamma}
\end{equation}
\begin{equation}
	\delta_\omega (r)=\frac{ -\xi^2 +\xi(\lambda+1-2e^\nu)-2(\lambda+1)(e^\nu-\lambda-1)+2\omega^2 r^2}{r e^\nu \xi(r)}
	\label{delta}
\end{equation}
\begin{equation}
	S_1=i\frac{4 Q\left\{2r e^\nu (e^\nu \mathcal{L}_F \bar{f}_{12})'+ 2(\lambda+1) (e^\nu \mathcal{L}_F\bar{f}_{12}) \right\}}{\omega r^3 \xi(r)}
	\label{S1}
\end{equation}
\begin{equation}
	S_2=\frac{4 Q\left\{2re^\nu (e^\nu \mathcal{L}_F \bar{f}_{12})'+  (\xi+2\lambda+2) (e^\nu \mathcal{L}_F\bar{f}_{12}) \right\}}{\omega r^2 e^\nu \xi(r)}.
	\label{S2}
\end{equation}
Here
\begin{eqnarray}
	\xi(r)&=& r e^\nu \nu' -2 e^\nu +l(l+1)  \nonumber \\
	&=& - r^2 (2\mathcal{L}+\Lambda)-2 r^{-2} Q^2\mathcal{L}_F -3e^\nu +2\lambda  +3 .
	\label{}
\end{eqnarray}
Equations\ (\ref{alpha}-\ref{S2}) are simplified using the background equations (\ref{EERaised1}) and  (\ref{EERaised2}).

We wish to combine Eqs.\ (\ref{dK}) and (\ref{dH1}) to a second order wave equation of the form
\begin{equation}
	\frac{d^2 R_{lm}^{(even)}}{dr_*} +[\omega^2-V^{(even)}(r)]R_{lm}^{(even)}=S_{lm}.
	\label{Rlm-eq}
\end{equation}
To do this, we follow the method outlined by Zerilli in \cite{Zerilli}.  The first step is to transform Eqs.\ (\ref{dK}) and (\ref{dH1}) to the form
\begin{equation}
	\frac{d\hat{K}}{d\hat{r}}= \hat{L} + \hat{S}_1
	\label{dKhat}
\end{equation}
\begin{equation}
	\frac{d\hat{L}}{d\hat{r}} = -[\omega^2 -V(\hat{r})] \hat{K} + \hat{S}_2,
	\label{dLhat}
\end{equation}
where the new variable $\hat{r}$ is given in terms of $r$ by $d\hat{r}/dr=1/n(r)$.
For brevity, one can rewrite Eqs.\ (\ref{dK}), (\ref{dH1}), (\ref{dKhat}) and (\ref{dLhat}) in the matrix form
\begin{equation}
	\frac{d\psi}{dr}=A \psi +S
	\label{matrix-psi}
\end{equation}
\begin{equation}
	\frac{d\hat{\psi}}{d\hat{r}} =\hat{A} \hat{\psi} + \hat{S},
	\label{matrix-psi-hat}
\end{equation}
where
\begin{equation}
	\psi=
	\left[ {\begin{array}{c}
			K\\
			\omega^{-1}H_1 \\
	\end{array} } \right], ~~
	A=
	\left[ {\begin{array}{cc}
			\alpha_\omega & \beta_\omega \\
			\gamma_\omega & \delta_\omega  \\
	\end{array} } \right],~~
	S=
	\left[ {\begin{array}{c}
			S_1 \\
			S_2  \\
	\end{array} } \right]
	\label{}
\end{equation}
and
\begin{equation}
	\hat{\psi}=
	\left[ {\begin{array}{c}
			\hat{K}\\
			\hat{L} \\
	\end{array} } \right], ~~
	\hat{A}=
	\left[ {\begin{array}{cc}
			0 & 1 \\
			-\omega^2+V& 0  \\
	\end{array} } \right],~~
	\hat{S}=
	\left[ {\begin{array}{c}
			\hat{S}_1 \\
			\hat{S}_2  \\
	\end{array} } \right].
	\label{}
\end{equation}
We now look for a transformation
\begin{equation}
	\psi=\mathcal{F} \hat{\psi},
	\label{F-transform}
\end{equation}
where
\begin{equation}
	\mathcal{F}=
	\left[ {\begin{array}{cc}
			f(r) & g(r)\\
			h(r) & k(r) \\
	\end{array} } \right]
	\label{}
\end{equation}
is to be determined.  Inserting Eq.\ (\ref{F-transform}) into (\ref{matrix-psi}) and then comparing the result to Eq.\ (\ref{matrix-psi-hat}) tells us that
\begin{equation}
	n\mathcal{F}^{-1}(A\mathcal{F}-\frac{d\mathcal{F}}{dr})=\hat{A}
	\label{}
\end{equation}
\begin{equation}
	\hat{S}= n\mathcal{F}^{-1}S.
	\label{Shat}
\end{equation}
Using the above equations, one can determine $n(r)$, $\mathcal{F}$, and consequently $\hat{S}$.
The results for the components of the matrix $\mathcal{F}$ are
\begin{eqnarray}
	f(r)=\frac{2 e^\nu \left(2 Q^2 \mathcal{L}_F+\lambda r^2\right)}{  r^3 \xi}+\frac{\lambda+1-e^\nu}{r}
	\label{fr}
\end{eqnarray}
\begin{equation}
	g(r)=1
	\label{gpr}
\end{equation}
\begin{eqnarray}
	h(r)=-i e^{-\nu} \left[ \frac{2 e^\nu \left(2 Q^2 \mathcal{L}_F+\lambda r^2\right)}{  r^2 \xi}+\lambda+1-e^\nu -\frac{\xi}{2}  \right]
	\label{hr}
\end{eqnarray}
\begin{equation}
	k(r)=-i r e^{-\nu},
	\label{}
\end{equation}
where we have used Eq.\ (\ref{EERaised2}) to simplify the above functions.
Also
\begin{equation}
	n(r)=e^{\nu},
	\label{nr}
\end{equation}
which shows that the new variable $\hat{r}$ is just the tortoise coordinate $r_*$.  Note that the functions $f(r)$ and $h(r)$ given in Eqs.\ (\ref{fr}) and (\ref{hr}) do not reduce to Zerilli's results in \cite{Zerilli} for the Reissner-Nordstr\"om case. For an explanation, see Appendix C. 

We can now express the potential in the following form 
\begin{eqnarray}
	V^{(even)}(r)&=&e^\nu \left\{\frac{8  e^\nu  Q^2 \mathcal{L}_F^2}{\xi \left( r^4 \mathcal{L}_F - Q^2 \mathcal{L}_{FF} \right)} +\frac{\xi}{r^2} +\frac{2}{r^2}(e^\nu -2\lambda-1)-\frac{4Q^2 \mathcal{L}_F}{r^4}  \right. \nonumber \\
	&&\left. ~~~~ -\frac{8}{r^4 \xi}(e^\nu -\lambda-1) (\lambda r^2+2 Q^2 \mathcal{L}_F) 
	-\frac{8 e^\nu}{r^6 \xi^2} (\lambda r^2+2 Q^2 \mathcal{L}_F)^2\right\} .
	\label{}
\end{eqnarray}
In addition, we can use Eq.\ (\ref{Shat}) to determine $\hat{S}_1$ and $\hat{S}_2$. 
Comparing Eqs.\ (\ref{dKhat}) and (\ref{dLhat}) with (\ref{Rlm-eq}), we get
\begin{equation}
	S_{lm}= \hat{S}_2+\frac{d\hat{S}_1}{dr_*}.
	\label{}
\end{equation}
It is also easy to combine Eqs.\ (\ref{dKhat}) and (\ref{F-transform}) to obtain
\begin{equation}
	K=f\hat{K}+g\hat{L}=fR_{lm}^{(even)}+\frac{dR_{lm}^{(even)}}{dr_*}-\hat{S}_1 .
	\label{}
\end{equation}
Using the results for $S_{lm}$ and $K$ we can write the final wave equations as
\begin{eqnarray}
	&&\hspace{-1cm}\frac{d^2R_{lm}^{(even)}}{dr_*^2}+[\omega^2-V^{(even)}(r)]R_{lm}^{(even)}=-\frac{16 i e^\nu Q \sqrt{\mathcal{L}_F}}{\omega r^5 \xi^2} 
	\left\{ \frac{1}{2} r^2 \xi (\xi-4\lambda-4) -4e^\nu Q^2 \mathcal{L}_{F} \right.  \nonumber \\
	&&~~~~~~~~~~~~~~~~~~~~~~~~~~~~~~~~~~~~~~~~~~~~~~\left. +2 r^2 e^\nu (\xi-\lambda) 
	-\frac{r^6 e^\nu \xi \mathcal{L}_F}{ r^4 \mathcal{L}_F - Q^2\mathcal{L}_{FF}}   \right\}  f_{lm}^{(even)}
	\label{WEeven1}
\end{eqnarray}
\begin{eqnarray}
	\frac{d^2f_{lm}^{(even)}}{dr_*^2}    +\left\{\omega^2 - e^\nu \left[ \frac{l(l+1) r^2 \mathcal{L}_F}{ r^4 \mathcal{L}_F - Q^2 \mathcal{L}_{FF}}    + \frac{e^{-\nu}}{4 \mathcal{L}_F^2}\left(3\left(\frac{d\mathcal{L}_F}{dr_*}\right)^2-2 \mathcal{L}_F \frac{d^2\mathcal{L}_F}{dr_*^2}\right)  \right. \right. \nonumber \\
	\left. \left. +  \frac{8 e^\nu Q^2 \mathcal{L}_F^2}{   \xi\left(r^4  \mathcal{L}_F - Q^2 \mathcal{L}_{FF} \right)} \right] \right\}f_{lm}^{(even)}
	\nonumber \\ 
	= - \frac{i \omega r^2 e^\nu Q  \mathcal{L}_F^{3/2}}{ r^4 \mathcal{L}_F - Q^2 \mathcal{L}_{FF} }  \left[f R_{lm}^{(even)}+ \frac{dR_{lm}^{(even)}}{dr_*}\right].
	\label{WEeven2}
\end{eqnarray}     
The equations (\ref{WEeven1}) and (\ref{WEeven2}) are similar in structure to the results found by Zerilli in \cite{Zerilli}.

\section{Stability for Even Parity Perturbations}
\label{Sec:SEven}

To make the wave equations more suitable for the stability analysis conducted in \cite{Moreno}, we want to eliminate the ${dR_{lm}^{(even)}}/{dr_*}$ term in Eq.\ (\ref{WEeven2}).  Below we explain how to systematically approach this problem.
We first rewrite Eqs.\ (\ref{f12-eq}), (\ref{dK}), and (\ref{dH1}) in the following form
\begin{equation}
	\frac{dK}{dr} = \alpha_\omega (r) K +\omega^{-1} \beta_\omega(r) H_1 + \omega^{-1} \varepsilon(r) F_1 +\omega^{-1} \eta(r) F 
	\label{dK-big}
\end{equation}
\begin{equation}
	\omega^{-1} \frac{dH_1}{dr}= \gamma_\omega (r) K +\omega^{-1} \delta_\omega(r) H_1 + \omega^{-1} \varkappa(r) F_1 +\omega^{-1} \rho(r) F
	\label{dH1-big}
\end{equation}
\begin{equation}
	\omega^{-1} \frac{dF_1}{dr}= \tau(r) K  + \omega^{-1} \phi(r) F_1 +\omega^{-1} \chi_\omega(r) F
	\label{dF1}
\end{equation}
\begin{equation}
	\frac{dF}{dr}= F_1,
	\label{dF}
\end{equation}
where $F = 2 e^\nu \mathcal{L}_F \bar{f}_{12}$ and 
\begin{equation}
	\varepsilon(r)=i\frac{4   e^\nu Q }{ r^2 \xi(r)}
	\label{}
\end{equation}
\begin{equation}
	\eta(r)=i\frac{4 Q (\lambda+1)  }{ r^3 \xi(r)}
	\label{}
\end{equation}
\begin{equation}
	\varkappa(r)=\frac{4   Q }{ r  \xi(r)}
	\label{}
\end{equation}
\begin{equation}
	\rho(r)=\frac{2 Q (\xi+2\lambda+2)   }{ r^2 e^\nu \xi(r)}
	\label{}
\end{equation}
\begin{equation}
	\tau(r)=-i \frac{2 r^2 Q \mathcal{L}_F^2    }{  e^\nu (r^4 \mathcal{L}_F - Q^2 \mathcal{L}_{FF})}
	\label{}
\end{equation}
\begin{equation}
	\phi(r)=- \nu' +\frac{\mathcal{L}_F'    }{  \mathcal{L}_F}
	\label{}
\end{equation}
\begin{equation}
	\chi_\omega(r)=-{e^{-2\nu}}\omega^2 + \frac{2 (\lambda+1) r^2 \mathcal{L}_F^2    }{  e^\nu (r^4 \mathcal{L}_F - Q^2 \mathcal{L}_{FF})}.
	\label{}
\end{equation}
We want to convert the system of equations (\ref{dK-big}-\ref{dF}) to
\begin{equation}
	\frac{d\mathcal{R}}{d\hat{r}} =\mathcal{R}_1
\end{equation}
\begin{equation}
	\frac{d\mathcal{R}_1}{d\hat{r}} =- [\omega^2 - V_{\mathcal{R}}(\hat{r})]\mathcal{R}+a(r) \mathcal{B}
\end{equation}
\begin{equation}
	\frac{d\mathcal{B}_1}{d\hat{r}} =b(r)\mathcal{R}- [\omega^2 - V_{\mathcal{B}}(\hat{r})] \mathcal{B}
\end{equation}
\begin{equation}
	\frac{d\mathcal{B}}{d\hat{r}} =\mathcal{B}_1,
\end{equation}
where the new variable $\hat{r}$ is given in terms of $r$ by $d\hat{r}/dr=1/n(r)$.  We first put the equations in matrix form:
\begin{equation}
	\frac{d\Psi}{dr}= M \Psi 
	\label{matrix-psi-large}
\end{equation}
\begin{equation}
	\frac{d\hat{\Psi}}{d\hat{r}} =\mathcal{M} \hat{\Psi} ,
	\label{Psi-hat-deq-large}
\end{equation}
where
\begin{equation}
	\Psi=
	\left[ {\begin{array}{c}
			K\\
			\omega^{-1}H_1 \\
			\omega^{-1}F_1 \\
			\omega^{-1}F\\
	\end{array} } \right], ~~
	M=
	\left[ {\begin{array}{cccc}
			\alpha_\omega & \beta_\omega & \varepsilon  & \eta  \\
			\gamma_\omega & \delta_\omega & \varkappa  & \rho  \\
			\tau  & 0 & \phi  & \chi_\omega \\
			0 & 0 & 1 & 0 \\
	\end{array} } \right]
	\label{}
\end{equation}
and
\begin{equation}
	\hat{\Psi}=
	\left[ {\begin{array}{c}
			\mathcal{R}\\
			\mathcal{R}_1 \\
			\mathcal{B}_1 \\
			\mathcal{B} \\
	\end{array} } \right], ~~
	\mathcal{M}=
	\left[ {\begin{array}{cccc}
			0 & 1 & 0 & 0 \\
			-\omega^2+V_\mathcal{R} & 0 & 0 & a \\
			b & 0 & 0 & 	-\omega^2+V_\mathcal{B} \\
			0 & 0 & 1 & 0 \\
	\end{array} } \right].
	\label{}
\end{equation}
We now look for a matrix transformation $\Psi=\mathcal{N} \hat{\Psi}$,
which combined with (\ref{matrix-psi-large}) gives
\begin{equation}
	n\mathcal{N}^{-1}(M\mathcal{N} -\frac{d\mathcal{N}}{dr}) = \mathcal{M}.
\end{equation}
We can now solve for $n$, $\mathcal{N}$, and $\mathcal{M}$.  We find $	n(r)= e^\nu$, which means $\hat{r}= r_*$. Putting these into $\Psi=\mathcal{N} \hat{\Psi}$ 
gives
\begin{equation}
	K=\frac{1}{\sqrt{2\lambda}(\lambda+1)}\mathcal{R}_1+\frac{1}{r\sqrt{2\lambda}(\lambda+1)}\left(\lambda+1-e^\nu+\frac{2\lambda e^\nu }{\xi}\right)\mathcal{R}-\frac{2 e^\nu Q\sqrt{\mathcal{L}_F}}{r^2\xi(\lambda+1)}\mathcal{B}
	\label{Keqn}
\end{equation}
\begin{equation}
	H_1=-\frac{i \omega r}{e^\nu \sqrt{2\lambda}(\lambda+1)}\mathcal{R}_1+\frac{i \omega}{e^\nu \sqrt{2\lambda}(\lambda+1)}\left(\frac{\xi}{2}-\lambda-1+e^\nu-\frac{2\lambda e^\nu }{\xi}\right)\mathcal{R}+\frac{2 i \omega  Q\sqrt{\mathcal{L}_F}}{r\xi(\lambda+1)}\mathcal{B}
	\label{Heqn}
\end{equation}
\begin{equation}
	\bar{f}_{12}=\frac{i \omega Q}{2 r e^\nu \sqrt{2\lambda}(\lambda+1)}\mathcal{R}+\frac{ i \omega }{4 e^\nu (\lambda+1) \sqrt{\mathcal{L}_F}}\mathcal{B}
	\label{f12eqn}
\end{equation}
and equation (\ref{Psi-hat-deq-large}) gives
\begin{eqnarray}
	&&\hspace{-1cm}\frac{d^2 \mathcal{R}}{dr_*^2}+\left\{\omega^2-e^\nu\left[ \frac{16\lambda e^\nu Q^2 {\mathcal{L}_F}}{r^4 \xi^2} +\frac{\xi}{r^2} - \frac{2(2\lambda +1 -e^\nu)}{r^2} +\frac{8\lambda (\lambda +1 -e^\nu)}{r^2 \xi}  +  \frac{8\lambda^2 e^\nu}{r^2 \xi^2}  \right]\right\}\mathcal{R}\nonumber \\
	&&= \sqrt{8\lambda \mathcal{L}_F}e^\nu Q 
	\left\{\frac{1}{r^3}  - \frac{4 e^\nu }{r^5 \xi^2} \left( \lambda r^2 +2 Q^2 \mathcal{L}_F \right) - \frac{4 }{r^3 \xi} \left( \lambda   +1 -e^\nu  \right)   -\frac{2r e^\nu \mathcal{L}_F }{\xi (r^4 \mathcal{L}_F-Q^2 \mathcal{L}_{FF})}   \right\}  \mathcal{B} \nonumber \\
	\label{we-even1}
\end{eqnarray}
\begin{eqnarray}
	&&\hspace{-1cm}\frac{d^2 \mathcal{B}}{dr_*^2}+\left\{\omega^2-e^\nu\left[ \frac{32 e^\nu Q^4 \mathcal{L}_F^2}{r^6 \xi^2} +\frac{16 \lambda e^\nu Q^2 \mathcal{L}_F}{r^4 \xi^2}  -  \frac{4 Q^2 \mathcal{L}_{F}}{ r^4  } \left(1-\frac{4(\lambda +1-e^\nu)}{\xi}  \right)  \right. \right. \nonumber \\
	&& \left.  \left.   -\frac{2(\lambda +1-e^\nu)-\xi}{r^2}  +  \frac{16 e^\nu Q^2 \mathcal{L}_{F}^2 }{  \xi (r^4 \mathcal{L}_{F}-Q^2 \mathcal{L}_{FF} )}  +\frac{r^2 \mathcal{L}_{F} (2\lambda +2 -e^\nu-\xi ) }{  r^4 \mathcal{L}_{F}-Q^2 \mathcal{L}_{FF}}    \right. \right. \nonumber \\
	&& \left.  \left.     -  \frac{7 r^6 e^\nu\mathcal{L}_{F}^2 }{  (r^4 \mathcal{L}_{F}-Q^2 \mathcal{L}_{FF} )^2}    +\frac{2 r^2 e^\nu \mathcal{L}_{F}^2 (3 r^8 \mathcal{L}_{F}-Q^4 \mathcal{L}_{FFF}) }{  (r^4 \mathcal{L}_{F}-Q^2 \mathcal{L}_{FF})^3}\right]\right\}\mathcal{B}\nonumber \\
	&&= \sqrt{8\lambda \mathcal{L}_F}e^\nu Q 
	\left\{\frac{1}{r^3}  - \frac{4 e^\nu }{r^5 \xi^2} \left( \lambda r^2 +2 Q^2 \mathcal{L}_F \right) - \frac{4 }{r^3 \xi} \left( \lambda   +1 -e^\nu  \right)   -\frac{2r e^\nu \mathcal{L}_F }{\xi (r^4 \mathcal{L}_F-Q^2 \mathcal{L}_{FF})}   \right\}  \mathcal{R} , \nonumber \\
	\label{we-even2}
\end{eqnarray}
where the relation between $\mathcal{R}$ and $\mathcal{B}$ and our original functions can easily be derived from equations (\ref{Keqn} - \ref{f12eqn}).  

The wave equations (\ref{we-even1}) and (\ref{we-even2}) can be rewritten in the form 
\begin{equation}
	e^{-\nu}\left[ \frac{1}{r}\frac{d}{dr_*}  \left(  r^2 \frac{d}{dr_*}  \left( \frac{\mathcal{R}}{r} \right)\right)  + \omega^2 \mathcal{R} \right] -V^{\text{II}}_{~11} \mathcal{R} - V_{~12}^{\text{II}} \mathcal{B} =0
	\label{}
\end{equation}
\begin{equation}
	e^{-\nu}\left[ \frac{1}{\sqrt{\mathcal{L}_F}} \frac{d}{dr_*} \left(  \mathcal{L}_F \frac{d}{dr_*}  \left( \frac{\mathcal{B}}{\sqrt{\mathcal{L}_F}} \right)  \right)+ \omega^2 \mathcal{B} \right] -V_{~22}^{\text{II}} \mathcal{B} - V_{~21}^{\text{II}} \mathcal{R} =0,
	\label{}
\end{equation}
where
\begin{equation}
	V_{~11}^{\text{II}}=- \frac{2\lambda }{r^2}   +\frac{8\lambda (\lambda +1 -e^\nu)}{r^2 \xi}  +  \frac{8\lambda^2 e^\nu}{r^2 \xi^2}  + \frac{16\lambda e^\nu Q^2 {\mathcal{L}_F}}{r^4 \xi^2} 
	\label{}
\end{equation}
\begin{equation}
	V_{~12}^{\text{II}}=V_{~21}^{\text{II}}= \sqrt{8\lambda \mathcal{L}_F} Q 
	\left\{\frac{1}{r^3}  - \frac{4 e^\nu }{r^5 \xi^2} \left( \lambda r^2 +2 Q^2 \mathcal{L}_F \right) - \frac{4 }{r^3 \xi} \left( \lambda   +1 -e^\nu  \right)   -\frac{2r e^\nu \mathcal{L}_F }{\xi (r^4 \mathcal{L}_F-Q^2 \mathcal{L}_{FF})}   \right\} 
	\label{}
\end{equation}
\begin{eqnarray}
	&&V_{~22}^{\text{II}}=\frac{32 e^\nu Q^4 \mathcal{L}_F^2}{r^6 \xi^2} +\frac{16 \lambda e^\nu Q^2 \mathcal{L}_F}{r^4 \xi^2}  -  \frac{4 Q^2 \mathcal{L}_{F}}{ r^4  } \left(1-\frac{4(\lambda +1-e^\nu)}{\xi}  \right)  \nonumber \\
	&&  -\frac{2(2\lambda +2-e^\nu -\xi)}{r^2}  +  \frac{16 e^\nu Q^2 \mathcal{L}_{F}^2 }{  \xi (r^4 \mathcal{L}_{F}-Q^2 \mathcal{L}_{FF} )}  +\frac{2 r^2 \mathcal{L}_{F} (2\lambda +2 +e^\nu-\xi ) }{  r^4 \mathcal{L}_{F}-Q^2 \mathcal{L}_{FF}}    \nonumber \\
	&&     -  \frac{16 r^6 e^\nu\mathcal{L}_{F}^2 }{  (r^4 \mathcal{L}_{F}-Q^2 \mathcal{L}_{FF} )^2}    +\frac{4 r^2 e^\nu \mathcal{L}_{F}^2 (3 r^8 \mathcal{L}_{F}-Q^4 \mathcal{L}_{FFF}) }{  (r^4 \mathcal{L}_{F}-Q^2 \mathcal{L}_{FF})^3}.
	\label{}
\end{eqnarray}
These equations agree with those in \cite{Moreno}.  

\section{Comparison with the Reissner-Nordstr\"om Results}
There are multiple mistakes in Eqs.\ (22-24) of \cite{Zerilli}.  These mistake are also noticed by Pani {\it et al.} in \cite{Pani}.  We provide the correct equations in (\ref{gravEven00}-\ref{gravEven22+33/2}).  In addition, Eqs.\ (\ref{fr}) and (\ref{hr}) for the Reissner-Nordstr\"om black hole, where $\mathcal{L}=F$, $\Lambda=0$, and $e^\nu=1-\frac{2M}{r}+\frac{q^2}{r^2}$, reduce to
\begin{equation}
	f(r)=\frac{16q^4-4q^2r [ 11M+2(\lambda -1)r ]  +r^2\{ 24M^2+12\lambda Mr+4\lambda (\lambda +1) r^2\} }{4r^3 (3Mr+\lambda r^2-2q^2)} 
	\label{RNf}
\end{equation} 
\begin{equation}
	h(r)=-i\left\{1-r^{-2}e^{-\nu}(Mr-q^2) -\frac{3Mr-4q^2}{3Mr+\lambda r^2-2q^2} \right\}.
	\label{RNh}
\end{equation} 
The above two functions are different than the $f(r)$ and $h(r)$ provided by Zerilli in \cite{Zerilli}.  However, they are in agreement with the results provided in \cite{Pani}.

The mistakes in \cite{Zerilli} appear to be typos, because Zerilli's final wave equations (Eqs.\ (48) and (49) of \cite{Zerilli}) are in good agreement with Eqs.\ (\ref{WEeven1}) and (\ref{WEeven2}) when reduced to the Reissner-Nordstr\"om case.  It is important, however,  to keep in mind that when Zerilli's wave equation for the even parity gravitational field (Eq.\ (48) of \cite{Zerilli}) is used, to avoid obtaining wrong results, Zerilli's function $f(r)$ should be replaced with the correct function provided in Eq.\ (\ref{RNf}) above.  


\def\jnl#1#2#3#4{{#1}{\bf #2} #3 (#4)}

\def\Zphys{{Z.\ Phys.} }
\def\jssc{{J.\ Solid State Chem.\ }}
\def\jpsJ{{J.\ Phys.\ Soc.\ Japan }}
\def\ptps{{Prog.\ Theoret.\ Phys.\ Suppl.\ }}
\def\PTP{{Prog.\ Theoret.\ Phys.\  }}
\def\LNC{{Lett.\ Nuovo.\ Cim.\  }}

\def\JMP{{J. Math.\ Phys.} }
\def\NPB{{Nucl.\ Phys.} B}
\def\NP{{Nucl.\ Phys.} }
\def\PLB{{Phys.\ Lett.} B}
\def\PL{{Phys.\ Lett.} }
\def\PRL{Phys.\ Rev.\ Lett.\ }
\def\PRA{{Phys.\ Rev.} A}
\def\PRB{{Phys.\ Rev.} B}
\def\PRD{{Phys.\ Rev.} D}
\def\PR{{Phys.\ Rev.} }
\def\PRe{{Phys.\ Rep.} }
\def\AP{{Ann.\ Phys.\ (N.Y.)} }
\def\RMP{{Rev.\ Mod.\ Phys.} }
\def\ZPC{{Z.\ Phys.} C}
\def\SCI{Science}
\def\CMP{Comm.\ Math.\ Phys. }
\def\MPLA{{Mod.\ Phys.\ Lett.} A}
\def\IJMPA{{Int.\ J.\ Mod.\ Phys.} A}
\def\IJMPB{{Int.\ J.\ Mod.\ Phys.} B}
\def\cmp{{Com.\ Math.\ Phys.}}
\def\JPA{{J.\  Phys.} A}
\def\CQG{Class.\ Quant.\ Grav.~}
\def\ATMP{Adv.\ Theoret.\ Math.\ Phys.~}
\def\AJP{Am.\ J.\ Phys.~}
\def\PRSA{{Proc.\ Roy.\ Soc.\ Lond.} A }
\def\ibid{{ibid.} }
\vskip 1cm

\leftline{\bf References}

\renewenvironment{thebibliography}[1]
        {\begin{list}{[$\,$\arabic{enumi}$\,$]}  
        {\usecounter{enumi}\setlength{\parsep}{0pt}
         \setlength{\itemsep}{0pt}  \renewcommand{\baselinestretch}{1.2}
         \settowidth
        {\labelwidth}{#1 ~ ~}\sloppy}}{\end{list}}


\end{document}